\documentclass[aip,apl,amsmath,amssymb,reprint,citeautoscript,longbibliography,groupedaddress]{revtex4-1}

\usepackage{graphicx}
\usepackage{dcolumn}
\usepackage{bm}

\usepackage[utf8]{inputenc}
\usepackage[T1]{fontenc}
\usepackage{mathptmx}
\usepackage{etoolbox}

\makeatletter
\def\@email#1#2{%
 \endgroup
 \patchcmd{\titleblock@produce}
  {\frontmatter@RRAPformat}
  {\frontmatter@RRAPformat{\produce@RRAP{*#1\href{mailto:#2}{#2}}}\frontmatter@RRAPformat}
  {}{}
}%
\makeatother

\begin{document}

\title[Electrical conductivity of a random nanowire network]{Electrical conductivity of a random nanowire network: \\ comparison of two-dimensional and quasi-three-dimensional models}
\author{Yuri Yu. Tarasevich}
 \email[Author to whom any correspondence should be addressed: ]{tarasevich@asu-edu.ru}
\affiliation{
Laboratory of Mathematical Modeling, Astrakhan Tatishchev State University, Astrakhan, Russia
}%
\author{Andrei V. Eserkepov}
 \email{dantealigjery49@gmail.com}
\affiliation{
Laboratory of Mathematical Modeling, Astrakhan Tatishchev State University, Astrakhan, Russia
}%
\date{\today}

\begin{abstract}
Although the two-dimensional model of random networks of metallic nanowires or carbon nanotubes is widely used, it significantly overestimates the number of contacts between elements compared to quasi-three-dimensional models. This, within the mean-field approximation, leads to overestimates of the electrical conductivity, especially when the main contribution to the system's electrical conductivity comes from the contact resistances between the conductors. In the two-dimensional model, the system's electrical conductivity depends quadratically on the conductor density, whereas in the three-dimensional model, this dependence is linear. We propose a simple modification of a two-dimensional model, which can capture the saturation effect of the number of contacts per conductor in a real nanowire network.
\end{abstract}

\maketitle

Random networks of metallic nanowires (NWs) and carbon nanotubes (CNTs) on an insulating substrate attract researchers' attention due to the use of such systems as components in various devices. A key issue is to identify the dependence of macroscopic physical properties of such networks (e.g., electrical conductivity) on the individual physical properties of their constituent elements (NWs, CNTs). While experimental studies reveal relationships between various macroscopic properties of the system as a whole, for example, the correlation between electrical conductivity and transparency of conducting films based on metallic NWs, it is very difficult, if at all possible, to experimentally obtain dependencies of macroscopic physical properties of the system on the properties of its constituent elements. To identify such dependencies, theoretical methods (e.g., dimensional analysis \cite{Benda2019}) and computer simulations \cite{Simoneau2013,Kim2018} are used.

Since, for both NWs and CNTs, the typical aspect ratio (i.e., length-to-width ratio) is on the order of several hundred (see, e.g., Refs.~\onlinecite{Ustinovich2020,Wang2026}), the model of one-dimensional conductors is widely used to simulate a random network of NWs or CNTs on an insulating substrate; i.e., it is assumed that conductors have zero width. The random network formed by such one-dimensional (1D) objects is considered two-dimensional (2D); a number of works are devoted to the study of the properties of such networks (see, e.g., Refs.~\onlinecite{Yi2004,Heitz2011,Kim2018}).

Let $N$ identical linear zero-width conductors of length $l$ be placed within a square domain $\mathcal{D}$ of size $L \times L$ ($L>l$) with periodic boundary conditions. The number density of conductors, i.e., the number of conductors per unit area, is
\begin{equation}\label{eq:numdens}
n = \frac{N}{ L^2}.
\end{equation}
The centers of the conductors are assumed to be independently and identically distributed within $\mathcal{D} \in \mathbb{R}^2$, i.e., $x,y \in [0;L]$, where $(x,y)$ are the coordinates of the center of the conductor under consideration, while the orientations of the conductors follow a given orientational probability distribution function $f_{\alpha}(\alpha)$.

In a recent work \cite{Tarasevich2026}, it was shown that, within the mean-field approach (MFA), the effective electrical conductivity of such a random network of conductors is
\begin{multline}\label{eq:MFAsigma-RC0}
\sigma  \left(R_\text{w} + R_\text{j}\right) \\=  n l^2 \frac{1 + \Delta}{\Delta} \left\langle \cos^2 \alpha \right\rangle
\left[ 1-\frac{2}{\Delta}\frac{\left(\lambda_1-1\right)\left(\lambda_1^{{ \langle N_\text{j} \rangle }+1}-1\right)}{\lambda_1\left(\lambda_1^{ \langle N_\text{j}\rangle  }+1\right)}\right].
\end{multline}
Here $\langle N_\text{j} \rangle$ is the mean number of contacts per conductor,
\begin{equation}\label{eq:Delta}
\Delta = \frac{R_\text{w}}{R_\text{j}},
\end{equation}
where $R_\text{w}$ is the electrical resistance of a conductor, and $R_\text{j}$ is the resistance of each junction (intersection) between any conductors;
\begin{equation}\label{eq:C2}
\left\langle \cos^2 \alpha \right\rangle = \int\limits_{-\pi/2}^{\pi/2} f_{\alpha}(\alpha)
  \cos^2 \alpha \, \mathrm{d}\alpha,
\end{equation}
\begin{equation}\label{eq:lambda1}
\lambda_1 = \frac{\mu + \sqrt{\mu^2 - 4}}{2},
\end{equation}
\begin{equation}\label{eq:mu}
 \mu = 2 + \frac{\Delta }{\left\langle N_\text{j} \right\rangle + 1}.
\end{equation}
In the case of uniformly distributed orientations
\begin{equation}\label{eq:cos2ODFs}
\left\langle \cos^2 \alpha \right\rangle = \frac{1}{2},
\end{equation}
and
\begin{equation}\label{eq:meanNj}
\langle N_\text{j} \rangle = \frac{2n l^2}{\pi}.
\end{equation}
In the case of large value of the number density of conductors and for $\Delta \gg 1$
\begin{equation}\label{eq:MFAsigma2DWDR}
\sigma  \left(R_\text{w} + R_\text{j}\right)=  \frac{n l^2}{2},
\end{equation}
and for $\Delta \ll 1$
\begin{equation}\label{eq:MFAsigma2DJDR}
\sigma  \left(R_\text{w} + R_\text{j}\right)=  \frac{n^2 l^4 }{12 \pi}
\end{equation}
(cf. Ref.~\onlinecite{Tarasevich2022}).

Although formula~\eqref{eq:MFAsigma-RC0} was derived for a network in which all conductors lie strictly in one plane, it is easy to show that it remains valid even when the conductors form a layer of finite but small thickness; it is only necessary that the inclination of all conductors relative to a certain plane be small, which corresponds to real networks.

An alternative approach is to use a quasi-three-dimensional model (Q3D) \cite{Simoneau2015,Forro2020,Daniels2021}. In this case, conductors are considered as thin but finite-diameter cylinders or spherocylinders. For instance, in a Q3D network proposed in Ref.~\onlinecite{Daniels2021}, each wire has a finite diameter and therefore occupies a certain volume in space. This volume is excluded for placing another wire. Wires are deposited sequentially, one by one; the vertical position of each new wire depends on the position of the wires already deposited. This arrangement results in some wires that are in contact in a perfectly 2D system being separated vertically in the Q3D system.

\begin{figure}[!htbp]
  \centering
  \includegraphics[width=\columnwidth]{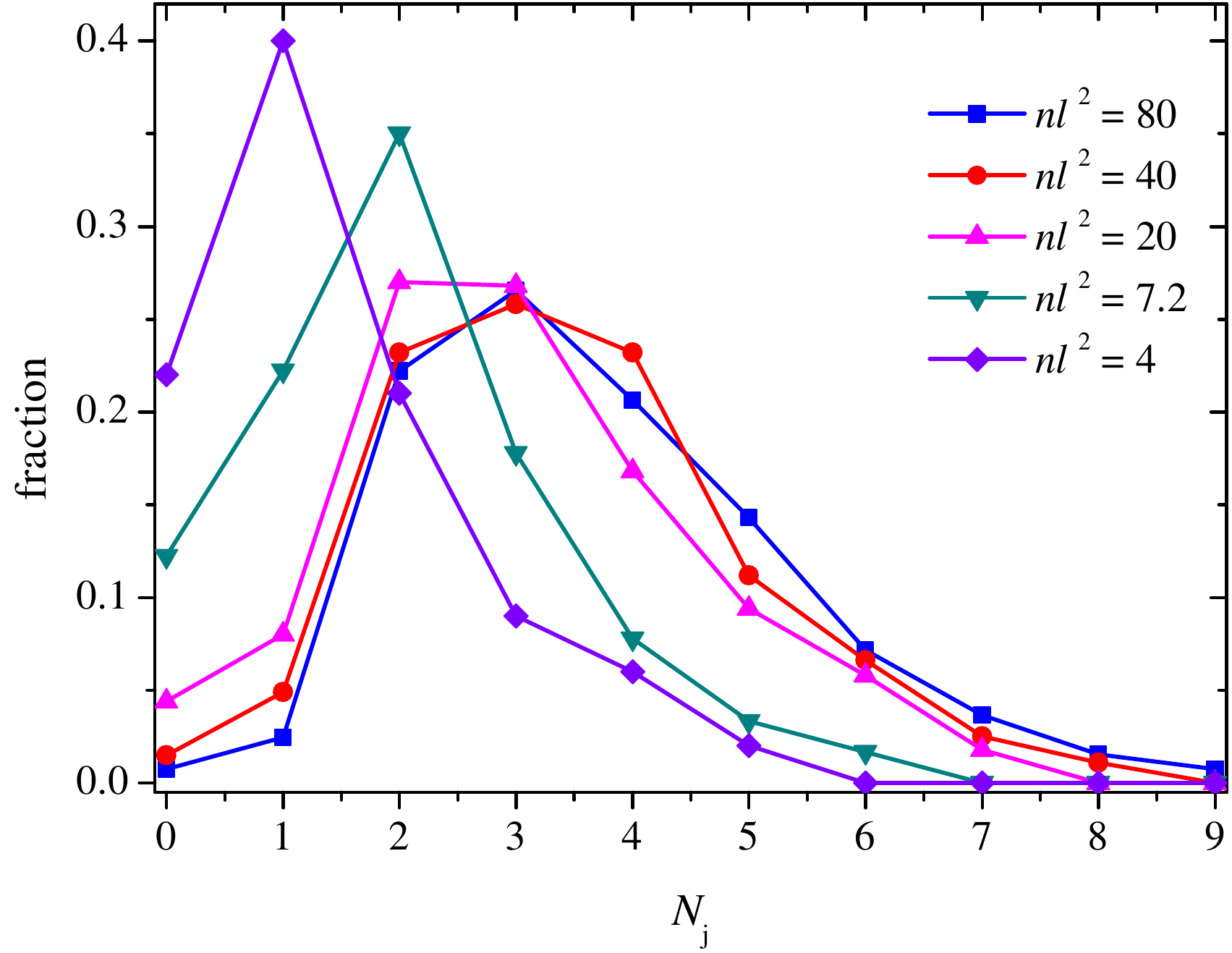}
  \caption{\label{fig:distrib}Distribution of the number of contacts per conductor for various values of the number density of conductors according to the data from Ref.~\onlinecite{Daniels2021}. }
\end{figure}

The properties of a network obtained by random deposition of such elongated objects of finite width onto a substrate are significantly different from the properties of a 2D network \cite{Daniels2021}. Figure~\ref{fig:distrib} shows the distribution of the number of contacts per conductor for various values of the number density of conductors according to computer simulation data for rigid cylindrical conductors with length $l = 6 \pm 3 \,\mu$m and diameter $d=20$~nm \cite{Daniels2021}. Thus, the aspect ratio was 300.

\begin{figure}[!htbp]
  \centering
  \includegraphics[width=\columnwidth]{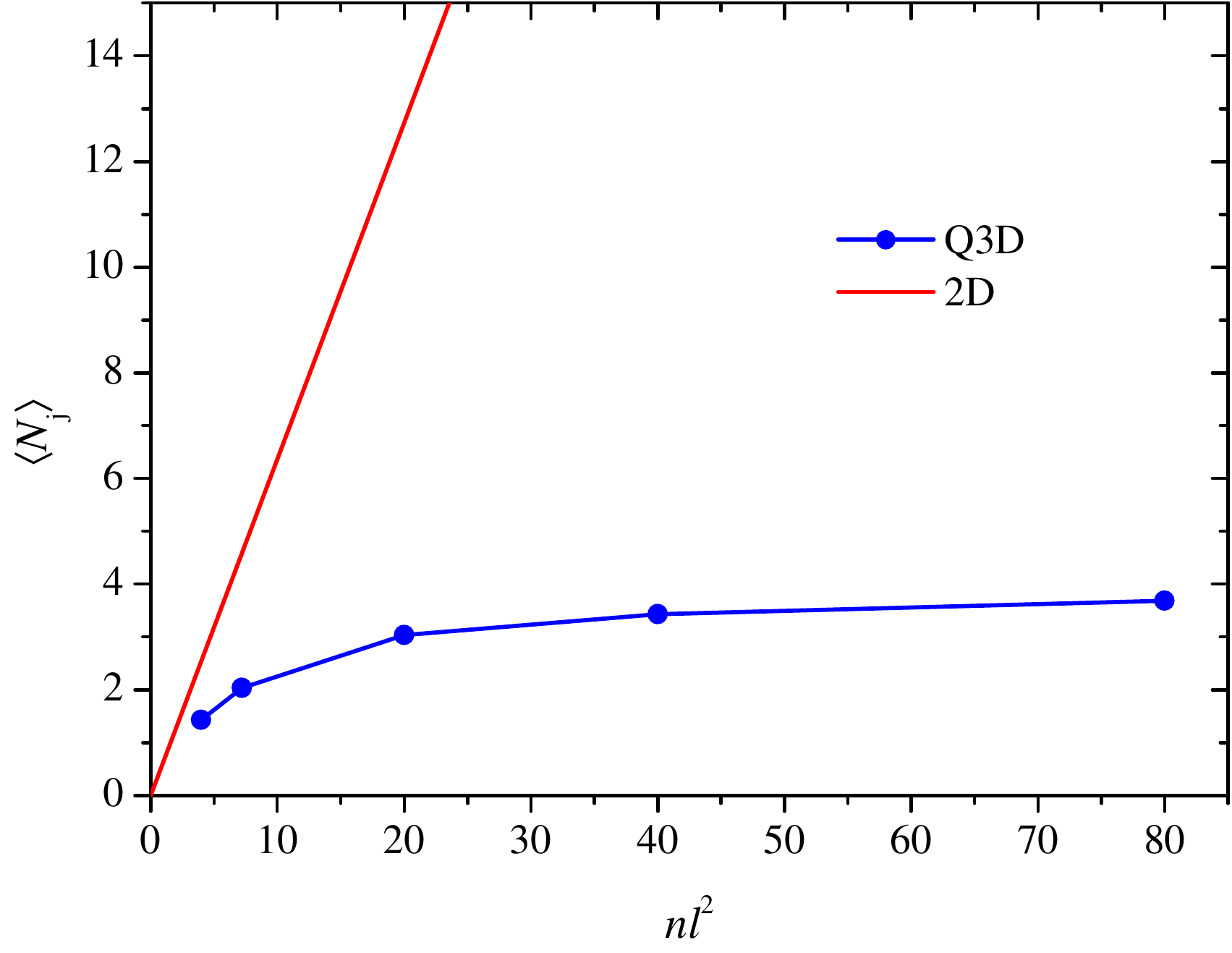}
  \caption{Dependencies of the mean number of contacts on the number density of conductors  for the 2D model~\eqref{eq:meanNj} and the Q3D model \cite{Daniels2021}. Conductor orientations are uniformly distributed.\label{fig:contacts}}
\end{figure}

Figure ~\ref{fig:contacts} shows the dependence of the average number of contacts per conductor on the number density of conductor for the 2D and Q3D models. For the 2D model, the number of contacts was calculated using~\eqref{eq:meanNj}; for the Q3D model, the values were taken from Ref.~\onlinecite{Daniels2021}, alternatively,  they can be obtained from the distributions presented in Fig.~\ref{fig:distrib}. While a linear increase is observed for a network of zero-width conductors, the dependence saturates for rigid conductors of finite width.

\begin{figure*}
  \centering
  \includegraphics[width=\textwidth]{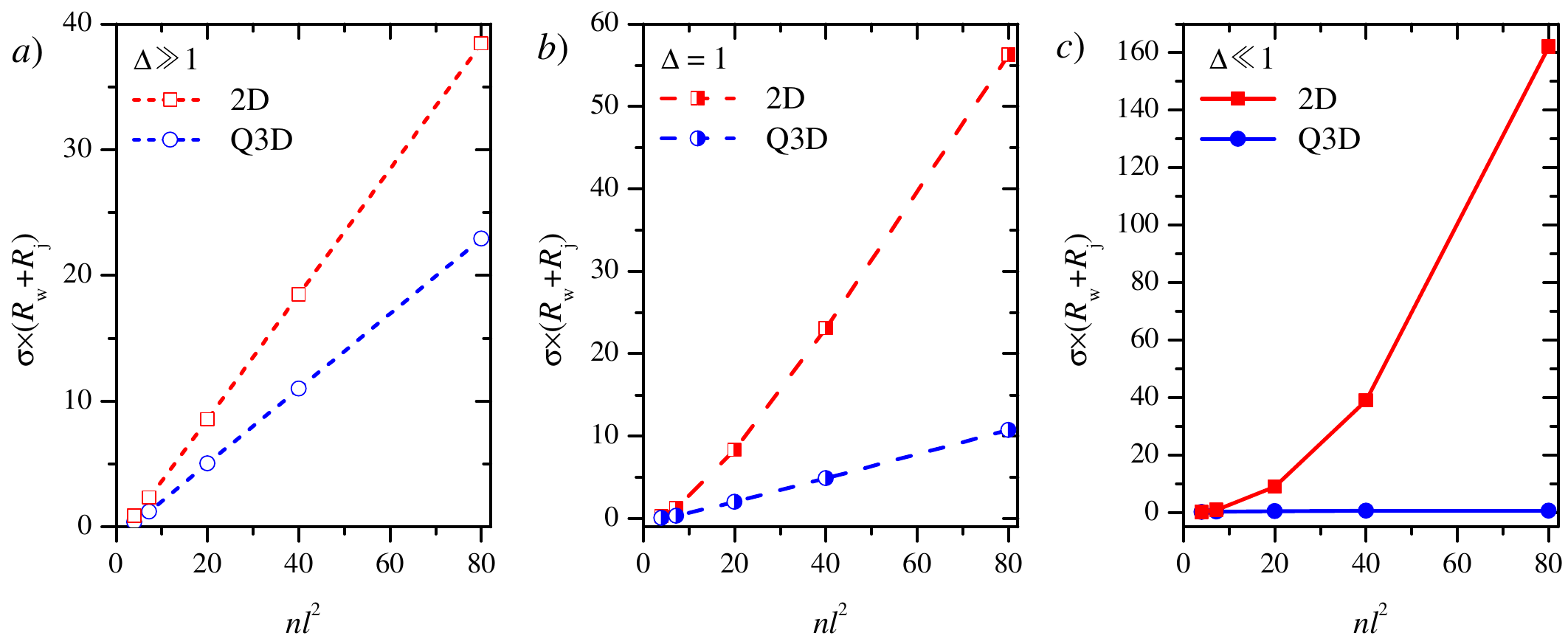}
  \caption{\label{fig:conductivity}Dependencies of electrical conductivity on the number density of conductors for the 2D and Q3D models, obtained within the MFA using formula~\eqref{eq:MFAsigma-RC0}. Conductor orientations are uniformly distributed. For the 2D model, the number of contacts per conductor was determined by formula~\eqref{eq:meanNj}, while  for the Q3D, it was extracted from Ref.~\onlinecite{Daniels2021}. }
\end{figure*}

Fundamentally different dependencies of the number of contacts per conductor on the number density of conductors in the 2D and Q3D models lead within the MFA to significantly different dependencies of electrical conductivity obtained on the number density (Fig.~\ref{fig:conductivity}). Dependencies were obtained within the MFA using formula~\eqref{eq:MFAsigma-RC0}. Conductor orientations are uniformly distributed. For the 2D model, the number of contacts per conductor was determined by formula~\eqref{eq:meanNj}, while  for the Q3D, it was extracted from Ref.~\onlinecite{Daniels2021}.  In the case where contact resistances dominate over conductor resistance, the differences in electrical conductivities obtained within the 2D and Q3D models are most significant and for the values of the number density presented in the figure reach two orders of magnitude. These differences decrease significantly with increasing parameter $\Delta$.

It is clear that both models are only limiting cases of infinitely soft conductors (2D model) and infinitely rigid conductors (Q3D model). Real conductors possess some flexibility, so the number of contacts per conductor, network connectivity, and its electrical conductivity should occupy some intermediate position between these two limiting cases.

More realistic and, accordingly, more complex models were proposed in Ref.~\onlinecite{Simoneau2015,Colasanti2016}. In Ref.~\onlinecite{Simoneau2015}, CNTs were modeled as spherocylinders with soft shells; besides straight spherocylinders, objects consisting of several segments were used to reproduce the waviness of CNTs.In Ref.~\onlinecite{Colasanti2016}, networks of CNTs were simulated. A CNT was modeled as a cylinder that could bend upon contact.

The authors of Ref.~\onlinecite{Colasanti2016} note that the 2D model overestimates the number of contacts between CNTs, even if the number of CNTs is very small. This artifact significantly impacts the reliability of the simulation, especially for dense films: since in the 2D model the network is more connected, there are more percolation paths; furthermore, since connections actually represent one of the main contributions to the film resistance, the 2D model cannot accurately reflect the real electrical behavior of the network.

In the case of a random three-dimensional (3D) packing of long rods, the average number of contacts per rod asymptotically in the limit of large length-to-width ratios ($>15$) tends to $\left\langle N_\text{j} \right\rangle/2 = 5.4 \pm 0.5$ \cite{Philipse1996,Philipse1996a,Williams2003}. This result was confirmed in the paper \cite{Blouwolff2006}, and in the case of the absence of compaction of the rod system the number of contacts is even less $\left\langle N_\text{j} \right\rangle/2 = 4.2 \pm 0.2$. The simulation results for the Q3D \cite{Daniels2021} case are quite consistent with those obtained for the three-dimensional packing of long rods, but the average number of contacts per rod is even smaller. Thus, it can be assumed that when considering nanowires as rigid, elongated objects with a large length-to-thickness ratio, the number of contacts quickly saturates with increasing nanowire concentration and equals several (up to 10) contacts per nanowire.

Assuming that with increasing number density of nanowires the number of contacts per nanowire quickly reaches saturation, we can estimate the behavior of electrical conductivity in this case. To do this, formulas~\eqref{eq:lambda1} and~\eqref{eq:mu} should be substituted into~\eqref{eq:MFAsigma-RC0}, after which it is necessary to find the limit~\eqref{eq:MFAsigma-RC0} as $\Delta \to 0$. When $\Delta \ll 1$, the formula~\eqref{eq:MFAsigma-RC0} becomes
\begin{equation}\label{eq:conductivityJDA}
\sigma \left(R_\text{w} + R_\text{j}\right) = \frac{n l^2}{12}
\frac{ \langle N_\text{j} \rangle^{2} - \langle N_\text{j} \rangle}{\langle N_\text{j} \rangle + 1} \left\langle \cos^2 \alpha \right\rangle.
\end{equation}
and, in the case of equally probable conductor orientations, it reduces to
\begin{equation}\label{eq:conductivityJDAiso}
\sigma \left(R_\text{w} + R_\text{j}\right) = \frac{n l^2}{24} \frac{ \langle N_\text{j} \rangle^{2} - \langle N_\text{j} \rangle}{\langle N_\text{j} \rangle + 1}.
\end{equation}
Thus, instead of the quadratic dependence of electrical conductivity on the nanowire concentration~\eqref{eq:MFAsigma2DJDR}, which is given by the 2D approximation, we have a linear dependence.

It is clear that the real situation is much more complex; the dependence of the number of contacts per  conductor on the number density of conductors significantly depends on our assumptions about the nature of interaction between conductor, the presence of a compressing (densifying) force, etc. If adhesion is excluded but gravity is considered significant, one can expect that the conductor will tend to occupy the lowest possible position, i.e., the supports will be located in hollows, leading to 4 contacts per newly added conductor. For example, the distributions presented in Ref.~\onlinecite{Daniels2021} for conductors with an aspect ratio of 300 indicate that the number of contacts can vary from 0 to 9 with a distribution maximum of approximately 3.

\begin{figure}[!htbp]
  \centering
  \includegraphics[width=\columnwidth]{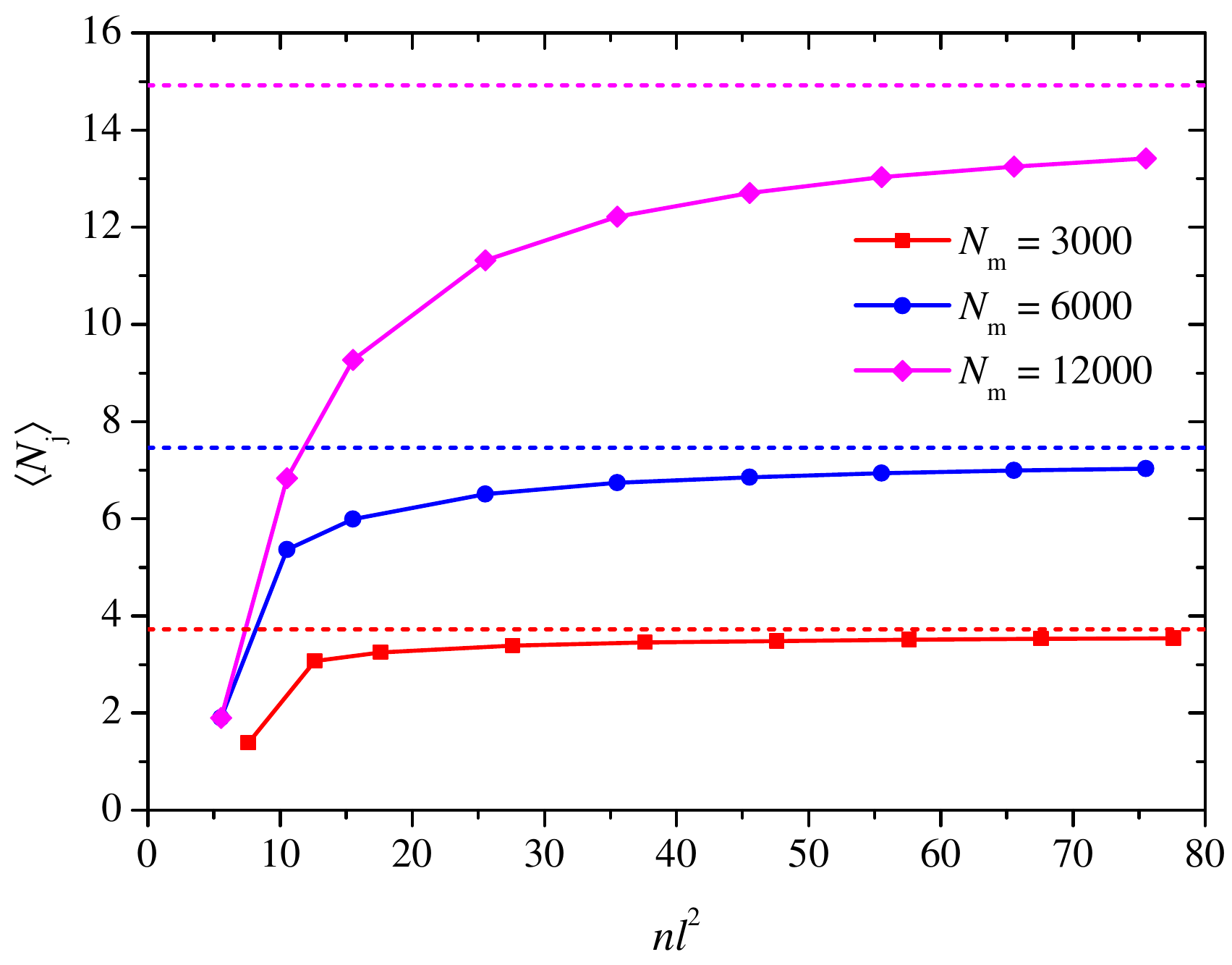}
  \caption{Dependencies of the mean number of contacts on the number density of conductors for the 2D model with memory for three values of the memory, $N_\text{m}$. Conductor orientations are uniformly distributed. Simulations were performed for $L=32$ and $l=1$ and averaged over 10 independent realizations. Standard error of the mean is of order of the marker size. The dashed lines correspond to the asymptotic values of $\langle N_\text{j} \rangle_\text{satur.}$  \eqref{eq:Njasymp}.\label{fig:NjperRod}}
\end{figure}

The saturation effect of the number of contacts per conductor in a real nanowire network can be simulated using a simple two-dimensional model with memory. Consider a square domain $\mathcal{D}$ on a plane. The domain size is $L \times L$. Periodic boundary conditions are used to reduce boundary effects. Segments of length $l$ are randomly and independently deposited into the domain $\mathcal{D}$. The segment orientations are equally probable. The segments are assigned ordinal numbers $i = 1\dots N$. When a new segment with number $j$ is deposited, its intersection with previously placed segments is checked. If segment $j$ intersects segment $i$, then contact (link) between them occurs only if the condition $j - i \leqslant N_\text{m}$ is satisfied, where $1\leqslant N_\text{m} < N$ plays the role of the memory depth. Within this model, the number of contacts per conductor saturates with increasing conductor concentration (Fig.~\ref{fig:NjperRod}). When $N_\text{m} \ll N$, the average number of contacts per rod tends to the value
\begin{equation}\label{eq:Njasymp}
  \langle N_\text{j} \rangle_\text{satur.} = \frac{4 l^2 N_m}{\pi L^2}.
\end{equation}

\begin{figure}[!htbp]
  \centering
  \includegraphics[width=\columnwidth]{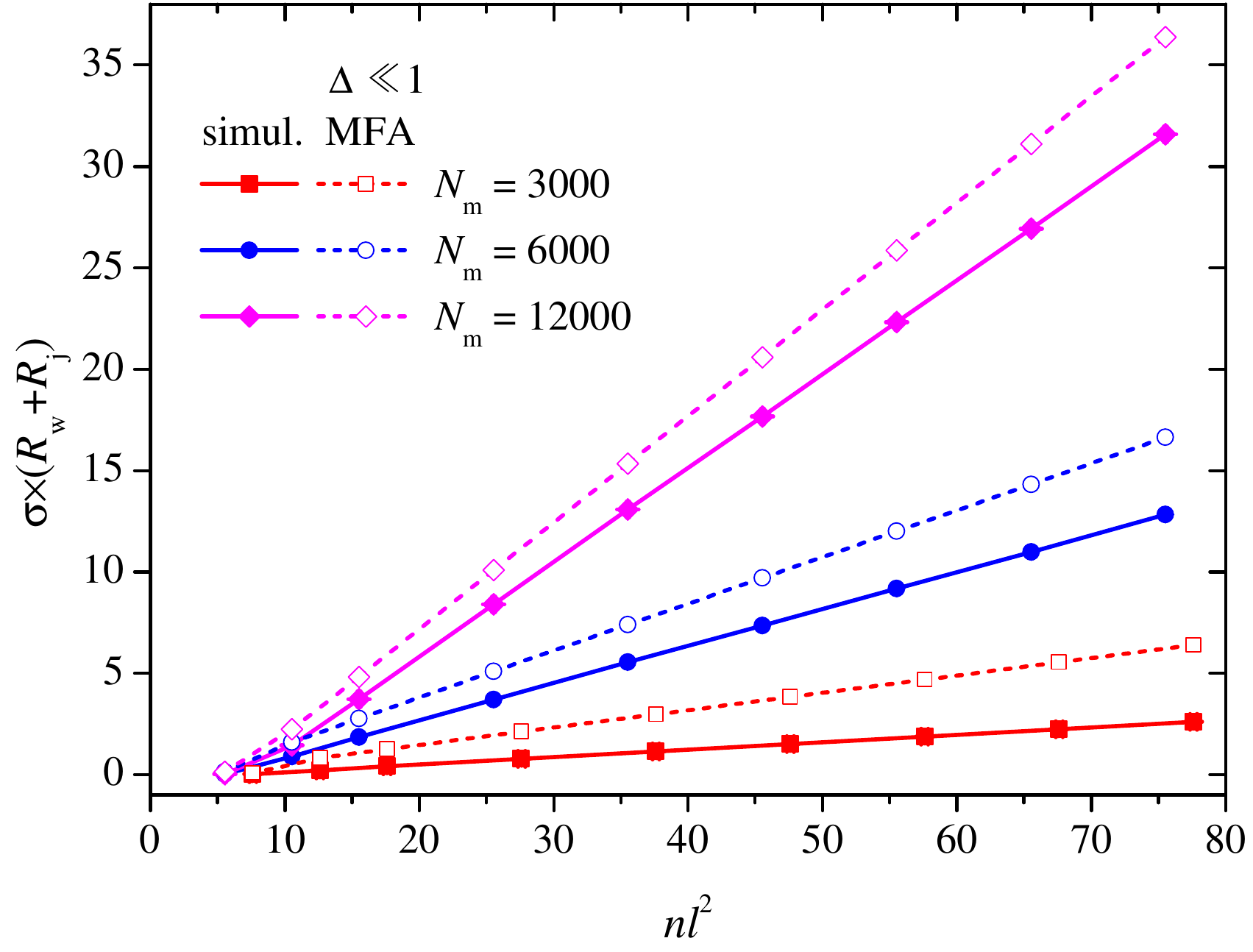}
  \caption{Dependencies of electrical conductivity on the number density of conductors for the 2D model with memory, obtained both within a mean-field-approach using formula~\eqref{eq:conductivityJDAiso} and by means a computer simulation. Conductor orientations are uniformly distributed. Simulations were performed for $L=32$, $l=1$, and $\Delta = 10^{-6}$ and averaged over 10 independent realizations. Standard error of the mean is of order of the marker size. \label{fig:conductivitymem}}
\end{figure}

Figure~\ref{fig:conductivitymem} compares the dependencies of the electrical conductivity on the number density of conductors obtained within the MFA and by simulation using the 2D model with memory for the case when the junction resistance dominates over the wire resistance ($\Delta \gg 1$). The reason for the overestimated electrical conductivity within the MFA was explained in Ref.~\onlinecite{Tarasevich2026}.

Extensive computer simulations using a 2D model were recently performed to explain experimental findings that the optoelectric performance of electrodes with cross-aligned metallic nanowires is superior to that of electrodes with randomly arranged nanowires~\cite{Grazioli2025}. Authors claimed that topology alone does not account for the advantages observed in cross-aligned arrangements, while the junction resistance is the key parameter influencing electrical conductivity. Moreover, Ref.~\onlinecite{Tarasevich2026} showed that, within the framework of a 2D model, the transition from a random arrangement of conductors to a cross-aligned one should lead to a decrease in electrical conductivity. The fact that in the case of randomly distributed nanowire networks the 2D model significantly overestimates the number of contacts provides an explanation for the contradiction between experimental observations and the predictions of mean field theory. Although accurate modeling of the properties of random nanowire networks requires knowledge of the mechanical properties of the nanowires and the details of the technological process of their deposition on the substrate, it is qualitatively clear that a network formed by linear thin conductors (finite-width rods) randomly deposited onto substrate and having uniformly distributed orientations has significantly lower connectivity than a similar network formed by zero-width rods. However, in the case of a two-layer network formed by cross-aligned conductors oriented in two mutually perpendicular directions, the differences between the 2D and Q3D models vanish.

\acknowledgments
Y.Y.T. thanks Avik Chatterjee for pointing out work Ref.~\onlinecite{Philipse1996} and Irina Vodolazskaya for careful reading of the manuscript and discussions.

\bibliography{Q3D}

\end{document}